\documentclass[graybox]{svmult}
\usepackage{mathptmx}       
\usepackage{helvet}         
\usepackage{courier}        
\usepackage{type1cm}        
\usepackage{makeidx}         
\usepackage{graphicx}        
\usepackage{float}	     
\usepackage{multicol}        
\usepackage[bottom]{footmisc}
\usepackage{amsmath} 
\usepackage{mathtools} 

\usepackage{epsfig}

\makeindex             

\begin{document} 

\title*{Structure and Width of the d$^\ast$(2380) Dibaryon} 
\author{Avraham Gal} 
\institute{Racah Institute of Physics, The Hebrew University, Jerusalem 91904, 
Israel \newline \email{avragal@savion.huji.ac.il}} 
\authorrunning{A.~Gal: the $d^\ast$(2380) dibaryon} 

\maketitle 

\abstract{In this contribution, dedicated to the memory of Walter Greiner, 
we discuss the structure and width of the recently established d$^\ast$(2380) 
dibaryon, confronting the consequences of our Pion Assisted Dibaryons 
hadronic model with those of quark motivated calculations. In particular, 
the relatively small width $\Gamma_{d^\ast}\approx 70$~MeV favors hadronic 
structure for the d$^\ast$(2380) dibaryon rather than a six-quark structure.}

\section{Walter Greiner: recollections} 
\label{sec:intro} 

\begin{figure}[!ht] 
\center{\includegraphics[width=\linewidth]{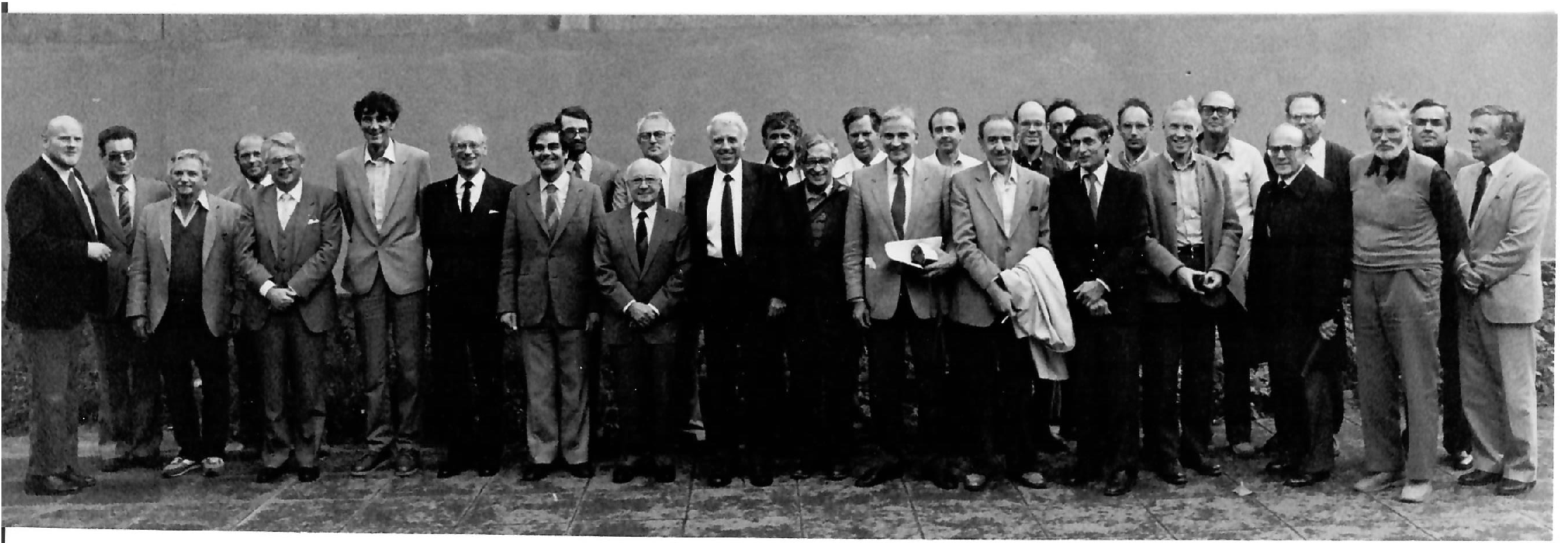}} 
\caption{Participants of the Frankfurt--Jerusalem Symposium in Frankfurt, 
1983. Walter Greiner is 3rd left on the 1st row. Eli Friedman is 5th right 
on the 2nd row. I'm missing in this photo.} 
\label{fig:greiner}
\end{figure}

This contribution is dedicated to the memory of Walter Greiner whose 
wide-ranging interests included exotic phases of matter. My first physics 
encounter with Walter was in Fall 1983 in a joint physics symposium hosted 
by him, see Fig.~\ref{fig:greiner}. 

\begin{figure}[htb] 
\begin{center} 
\includegraphics[width=.48\linewidth]{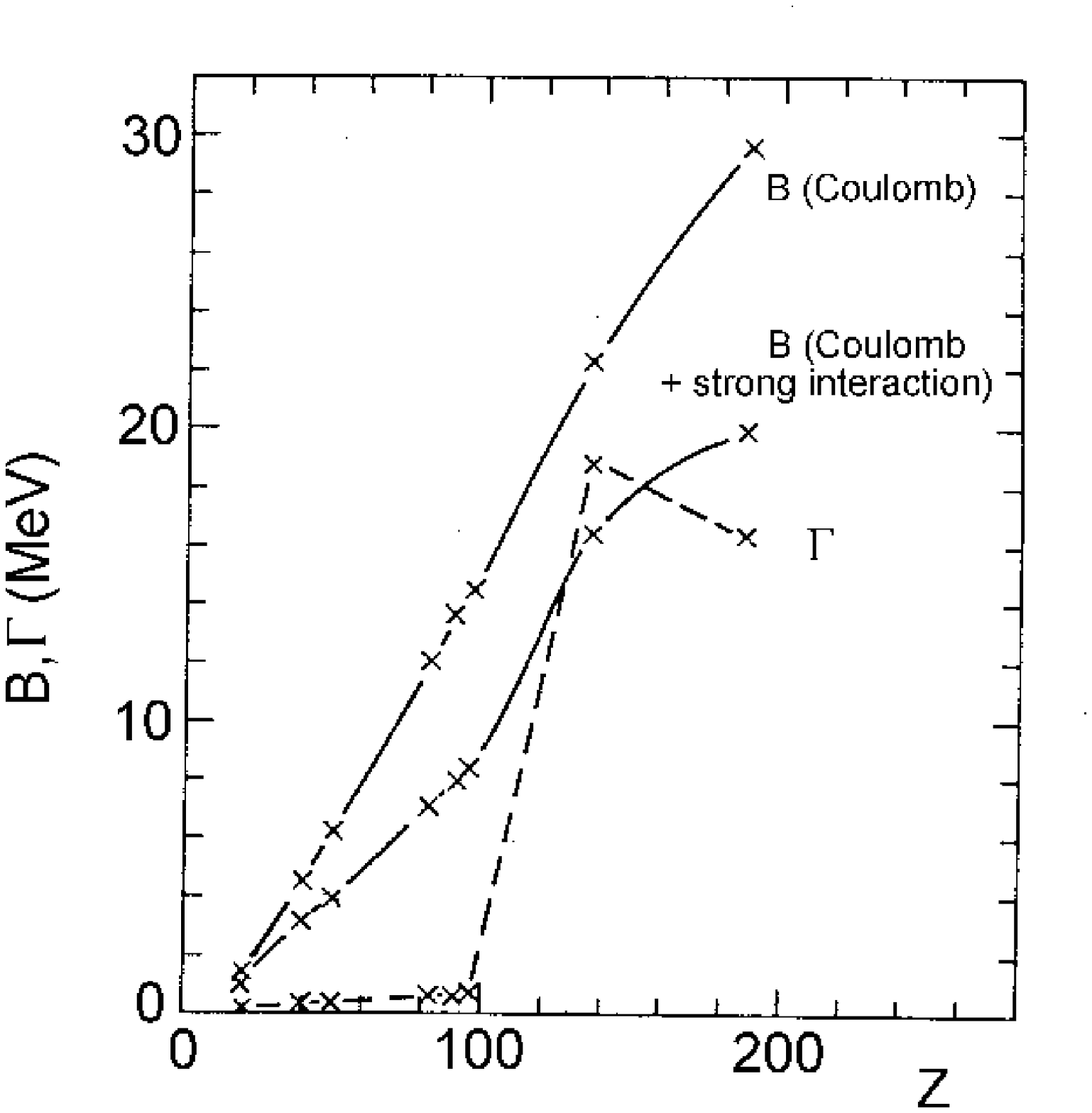} 
\includegraphics[width=.48\linewidth,height=5.5cm]{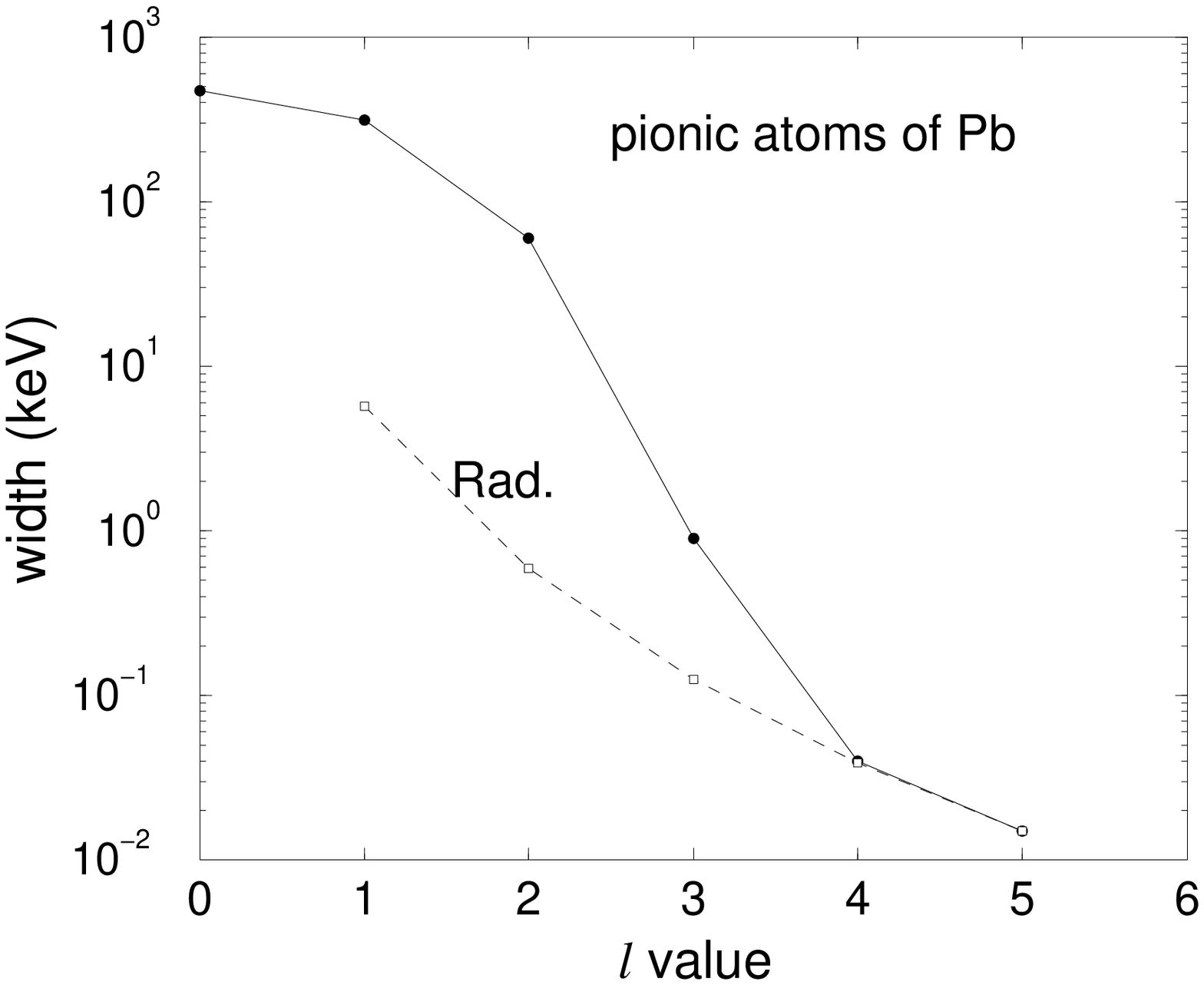} 
\caption{First prediction of deeply bound pionic atom states \cite{FS85}. 
The left panel shows binding energies and widths in 1s deeply bound pionic 
states~\cite{FS85}, and the right panel shows the width saturation in circular 
states of pionic atoms of Pb~\cite{FG07}.} 
\label{fig:FS85} 
\end{center} 
\end{figure}

Greiner's wide-ranging interests included also superheavy elements, so it was 
quite natural for him to ask my good colleague Eli Friedman, a leading figure 
in exotic atoms, whether extrapolating pionic atoms to superheavy elements 
would shed light on a then-speculated pion condensation phase. Subsequently 
in 1984 Eli spent one month in Frankfurt at Greiner's invitation, concluding 
together with Gerhard Soff~\cite{FS85} that the strong-interaction 
$\pi^-_{1s}$ repulsive energy shift known from light pionic atoms persists 
also in superheavy elements, as shown in Fig.~\ref{fig:FS85}-left, thereby 
ruling out pion condensation for large $Z$. However, quite surprisingly, they 
also found that 1s \& 2p $\pi^-$ atomic states in normal heavy elements up to 
$Z\approx 100$ have abnormally small widths of less than 1~MeV owing to the 
repulsive $\pi$-nucleus strong interaction within the nuclear volume. Hence 
`deeply bound' states (DBS) in pionic atoms are experimentally resolvable, 
although they cannot be populated radiatively as in light pionic atoms because 
the absorption width in the higher 3d state exceeds its radiative width by 
almost two orders of magnitude, as shown in Fig.~\ref{fig:FS85}-right. 

Friedman and Soff's 1985 prediction of DBS was repeated three years later 
by Toki and Yamazaki~\cite{TY88}, who apparently were not aware of it, and 
verified experimentally in 1996 at GSI in a (d,$\,^3$He) reaction on $^{208}
$Pb~\cite{Yam96}. Subsequent experiments on Pb and Sn isotopes have yielded 
accurate data on several other pionic-atom DBS~\cite{yama12}, showing clear 
evidence in support of Weise's 1990 conjecture of partial chiral symmetry 
restoration in the nuclear medium due to a renormalized isovector $s$-wave 
$\pi N$ interaction through the decrease of the pion-decay constant $f_{\pi}
$~\cite{weise90}. However, the few DBS established so far are still short 
of providing on their own the precision reached by comprehensive fits to 
{\it all} (of order 100) pionic atom data, dominantly in higher atomic orbits, 
in substantiating this conjecture; for a recent review on the state of the art 
in pionic atoms see Ref.~\cite{FG14}. 

My own encounters with Walter Greiner and several of his colleagues in 
Frankfurt during several Humboldt-Prize periods in the 1990s focused on 
developing the concept of Strange Hadronic Matter~\cite{SHM93,SHM94,SHM00} 
and also on studying Kaon Condensation~\cite{KC94}. I recall fondly that 
period. Here I highlight another exotic phase of matter: non-strange 
Pion Assisted Dibaryons, reviewed by me recently in Ref.~\cite{gal16}.

\section{Pion assisted $N\Delta$ and $\Delta\Delta$ dibaryons} 

\subsection{The Dyson-Xuong 1964 prediction} 

Non-strange $s$-wave dibaryon resonances ${\cal D}_{IS}$ with isospin $I$ 
and spin $S$ were predicted by Dyson and Xuong in 1964~\cite{DX64} as early 
as SU(6) symmetry proved successful, placing the nucleon $N(939)$ and its 
$P_{33}$ $\pi N$ resonance $\Delta(1232)$ in the same ${\bf 56}$ multiplet. 
These authors chose the ${\bf 490}$ lowest-dimension SU(6) multiplet in the 
$\bf{56\times 56}$ direct product containing the flavor-SU(3) $\overline{\bf 
10}$ and ${\bf 27}$ multiplets in which the deuteron, ${\cal D}_{01}$, and 
$NN$ virtual state, ${\cal D}_{10}$, are classified. Four more non-strange 
dibaryons emerged in this scheme, with masses listed in Table~\ref{tab:dyson} 
in terms of constants $A$ and $B$. Identifying $A$ with the $NN$ threshold 
mass 1878~MeV, the value $B\approx 47$~MeV was derived by assigning 
${\cal D}_{12}$ to the $pp\leftrightarrow \pi^+ d$ coupled-channel resonance 
behavior noted then at 2160~MeV, near the $N\Delta$ threshold (2.171~MeV). 
This led in particular to a predicted mass $M=2350$~MeV for the $\Delta\Delta$ 
dibaryon candidate ${\cal D}_{03}$ assigned at present to the recently 
established d$^\ast$(2380) resonance~\cite{clement17}. 

\begin{table}[hbt] 
\begin{center}
\caption{Predicted masses of non-strange $L=0$ dibaryons ${\cal D}_{IS}$ 
with isospin $I$ and spin $S$, using the Dyson-Xuong SU(6)$\to$SU(4) mass 
formula $M=A+B[I(I+1)+S(S+1)-2]$~\cite{DX64}.} 
\begin{tabular}{ccccccccccccc}
\hline
${\cal D}_{IS}$ & & ${\cal D}_{01}$ & & ${\cal D}_{10}$ & & ${\cal D}_{12}$ 
& & ${\cal D}_{21}$ & & ${\cal D}_{03}$ & & ${\cal D}_{30}$ \\
\hline
$BB'$ & & $NN$ & & $NN$ & & $N\Delta$ & & $N\Delta$ & & $\Delta\Delta$ & & 
$\Delta\Delta$ \\
SU(3)$_{\rm f}$ & & $\overline{\bf 10}$ & & ${\bf 27}$ & & ${\bf 27}$ & & 
${\bf 35}$ & & $\overline{\bf 10}$ & & ${\bf 28}$ \\
$M({\cal D}_{IS})$ & & $A$ & & $A$ & & $A+6B$ & & $A+6B$ & & $A+10B$ & & 
$A+10B$ \\
\hline
\end{tabular}
\label{tab:dyson}
\end{center}
\end{table}

In retrospect, the choice of the ${\bf 490}$ lowest-dimension SU(6) multiplet, 
with Young tableau denoted [3,3,0], is not accidental. This [3,3,0] is the one 
adjoint to [2,2,2] for color-SU(3) singlet six-quark (6q) state, thereby 
ensuring a totally antisymmetric color-flavor-spin-space 6q wavefunction, 
assuming a totally symmetric $L=0$ orbital component. For non-strange 
dibaryons, flavor-SU(3) reduces to isospin-SU(2), whence flavor-spin SU(6) 
reduces to isospin-spin SU(4) in which the [3,3,0] representation corresponds 
to a ${\bf 50}$ dimensional representation consisting of precisely the $I,S$ 
values of the dibaryon candidates listed in Table~\ref{tab:dyson}, as also 
noted recently in Ref.~\cite{PPL15}. Since the ${\bf 27}$ and $\overline{
\bf 10}$ flavor-SU(3) multiplets accommodate $NN$ $s$-wave states that are 
close to binding ($^1S_0$) or weakly bound ($^3S_1$), we focus here on the 
${\cal D}_{12}$ and ${\cal D}_{03}$ dibaryon candidates assigned to these 
flavor-SU(3) multiplets.

\subsection{Pion assisted dibaryons} 

The pion plays a major role as a virtual particle in binding or almost binding 
$NN$ $s$-wave states. The pion as a real particle interacts strongly with 
nucleons, giving rise to the $\pi N$ $P_{33}$ $p$-wave $\Delta$(1232) 
resonance. Can it also assist in binding two nucleons into $s$-wave $N\Delta$ 
states? And once we have such $N\Delta$ states, can the pion assist in binding 
them into $s$-wave $\Delta\Delta$ states? This is the idea behind the concept 
developed by Garcilazo and me of pion assisted dibaryons~\cite{GG13,GG14}, or 
more generally meson assisted dibaryons to go beyond the non-strange sector, 
see Ref.~\cite{gal16} for review. 

As discussed in the next subsection, describing $N\Delta$ systems in terms of 
a stable nucleon ($N$) and a two-body $\pi N$ resonance ($\Delta$) leads to 
a well defined $\pi NN$ three-body model in which $IJ=12$ and $21$ resonances 
identified with the ${\cal D}_{12}$ and ${\cal D}_{21}$ dibaryons of 
Table~\ref{tab:dyson} are generated. This relationship between $N\Delta$ and 
$\pi NN$ may be generalized into relationship between a two-body $B\Delta$ 
system and a three-body $\pi NB$ system, where the baryon $B$ stands for $N, 
\Delta, Y$ (hyperon) etc. In order to stay within a three-body formulation 
one needs to assume that the baryon $B$ is stable. For $B=N$, this 
formulation relates the $N\Delta$ system to the three-body $\pi NN$ system. 
For $B=\Delta$, once properly formulated, it relates the 
$\Delta\Delta$ system to the three-body $\pi N\Delta$ system, suggesting to 
seek $\Delta\Delta$ dibaryon resonances by solving $\pi N\Delta$ Faddeev 
equations, with a stable $\Delta$. The decay width of the $\Delta$ resonance 
is considered then at the penultimate stage of the calculation. In terms 
of two-body isobars we have then a coupled-channel problem $B\Delta
\leftrightarrow\pi D$, where $D$ stands generically for appropriate dibaryon 
isobars: (i) ${\cal D}_{01}$ and ${\cal D}_{10}$, which are the $NN$ isobars 
identified with the deuteron and virtual state respectively, for $B=N$, and 
(ii) ${\cal D}_{12}$ and ${\cal D}_{21}$ for $B=\Delta$. 

\begin{figure}[hbt] 
\begin{center} 
\includegraphics[width=0.8\textwidth]{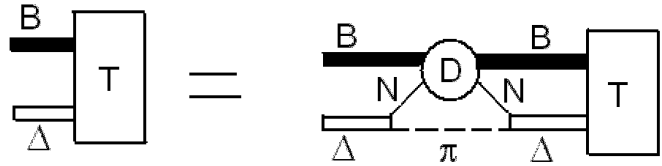} 
\caption{Diagrammatic representation of the integral equation for the 
$B\Delta$ $T$ matrix, derived by using separable pairwise interactions in 
$\pi NB$ Faddeev equations~\cite{GG14} and solved numerically to calculate 
$B\Delta$ dibaryon resonance poles for $B=N,\,\Delta$.} 
\label{fig:piNB} 
\end{center} 
\end{figure} 

Within this model, and using separable pairwise interactions, 
the coupled-channel $B\Delta -\pi D$ eigenvalue problem reduces to a single 
integral equation for the $B\Delta$ $T$ matrix shown diagrammatically in 
Fig.~\ref{fig:piNB}, where starting with a $B\Delta$ configuration the 
$\Delta$-resonance isobar decays into $\pi N$, followed by $NB\to NB$ 
scattering through the $D$-isobar with a spectator pion, and ultimately 
by means of the inverse decay $\pi N\to\Delta$ back into the $B\Delta$ 
configuration. We note that the interaction between the $\pi$ meson and 
$B$ is neglected for $B=\Delta$, for lack of known $\pi\Delta$ isobar 
resonances in the relevant energy range.

\subsection{$N\Delta$ dibaryons} 

The ${\cal D}_{12}$ dibaryon of Table~\ref{tab:dyson} shows up clearly in 
the Argand diagram of the $NN$ $^1D_2$ partial wave which is coupled above 
the $NN\pi$ threshold to the $I=1$ $s$-wave $N\Delta$ channel. Values of 
${\cal D}_{12}$ and ${\cal D}_{21}$ pole positions $W=M-{\rm i}\Gamma/2$ from 
our hadronic-model three-body $\pi NN$ Faddeev calculations~\cite{GG13,GG14} 
described in the previous subsection are listed in Table~\ref{tab:NDel} 
together with results of phenomenological studies that include (i) early $NN$ 
phase shift analyses \cite{arndt87} and (ii) $pp \leftrightarrow np\pi^+$ 
coupled-channels analyses~\cite{hosh92}. The ${\cal D}_{12}$ mass and width 
values calculated in the Faddeev hadronic model version using $r_{\Delta}
\approx\,1.3$~fm are remarkably close to the phenomenologically derived ones, 
whereas the mass evaluated in the version using $r_{\Delta}\approx\,0.9$~fm 
agrees with that assumed in the Dyson-Xuong pioneering discussion~\cite{DX64}. 

\begin{table}[hbt] 
\begin{center} 
\caption{${\cal D}_{12}$ and ${\cal D}_{21}$ $N\Delta$ dibaryon $S$-matrix 
pole positions $W=M-{\rm i}\frac{\Gamma}{2}$ (in MeV), obtained by solving 
the $N\Delta$ $T$-matrix integral equation of Fig.~\ref{fig:piNB}~\cite{GG14}, 
are listed for two choices of the $\pi N$ $P_{33}$ form factor specified by 
a radius parameter $r_{\Delta}$ (in fm) together with two phenomenological 
values. The last column lists the results of a nonrelativistic meson-exchange 
Faddeev calculation.} 
\begin{tabular}{ccccccccccc} 
\hline 
$N\Delta$ & & \multicolumn{3}{c}{Phenomenological} & & \multicolumn{3}{c}
{Faddeev (present)} & & \multicolumn{1}{c}{Faddeev (non rel.)} \\ 
${\cal D}_{IS}$ & & Ref.~\cite{arndt87} & & Ref.~\cite{hosh92} & & $r_{
\Delta}\approx\,1.3$ & & $r_{\Delta}\approx\,0.9$ & & Ref.~\cite{ueda82} \\  
\hline 
${\cal D}_{12}$ & & 2148$-{\rm i}$63 & & 2144$-{\rm i}$55 & & 2147$-{\rm i}$60 
& & 2159$-{\rm i}$70 & & 2116$-{\rm i}$61  \\
${\cal D}_{21}$ & &--& &--& & 2165$-{\rm i}$64 & & 2169$-{\rm i}$69 & &-- \\ 
\hline 
\end{tabular} 
\label{tab:NDel} 
\end{center} 
\end{table} 

Recent $pp\to pp\pi^+\pi^-$ production data~\cite{wasa18} locate the 
${\cal D}_{21}$ dibaryon resonance almost degenerate with the ${\cal D}_{12}$. 
Our $\pi NN$ Faddeev calculations produce it about 10-20~MeV higher than the 
${\cal D}_{12}$, see Table~\ref{tab:NDel}. The widths of these near-threshold 
$N\Delta$ dibaryons are, naturally, close to that of the $\Delta$ resonance. 
We note that only $^3S_1$ $NN$ enters the calculation of the ${\cal D}_{12}$ 
resonance, while for the ${\cal D}_{21}$ resonance calculation only $^1S_0$ 
$NN$ enters, both with maximal strength. Obviously, with the $^1S_0$ 
interaction the weaker of the two, one expects indeed that the ${\cal D}_{21}$ 
resonance lies above the ${\cal D}_{12}$ resonance. Moreover, these two 
dibaryon resonances differ also in their flavor-SU(3) classification, see 
Table~\ref{tab:dyson}, which is likely to push up the ${\cal D}_{21}$ further 
away from the ${\cal D}_{12}$. Finally, the $N\Delta$ $s$-wave states with 
$IJ=$ $11$ and $22$ are found not to resonate in the $\pi NN$ Faddeev 
calculations~\cite{GG14}.

\subsection{$\Delta\Delta$ dibaryons} 

\begin{figure}[!ht] 
\begin{center} 
\includegraphics[width=0.48\textwidth,height=5cm]{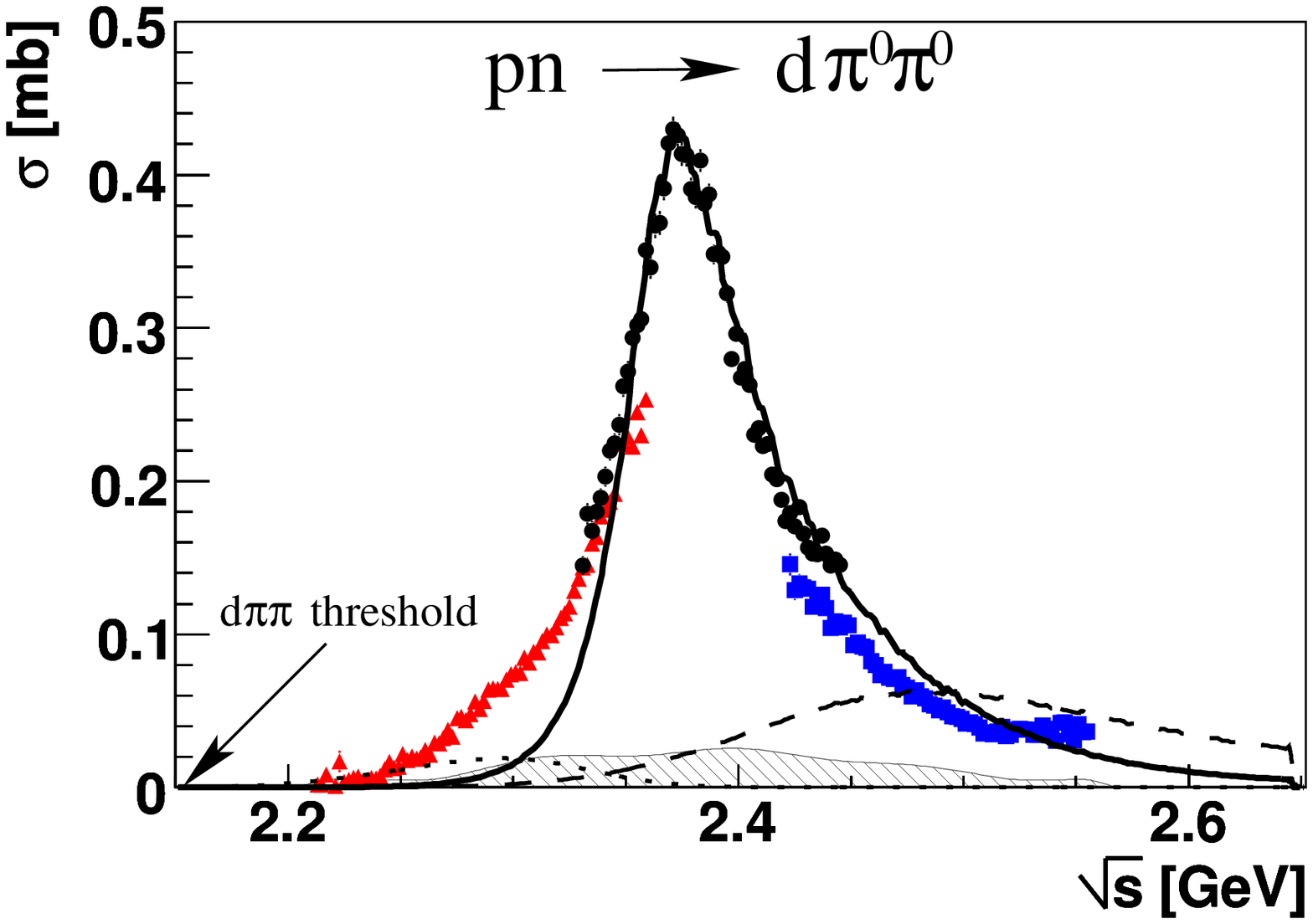} 
\includegraphics[width=0.48\textwidth,height=5cm]{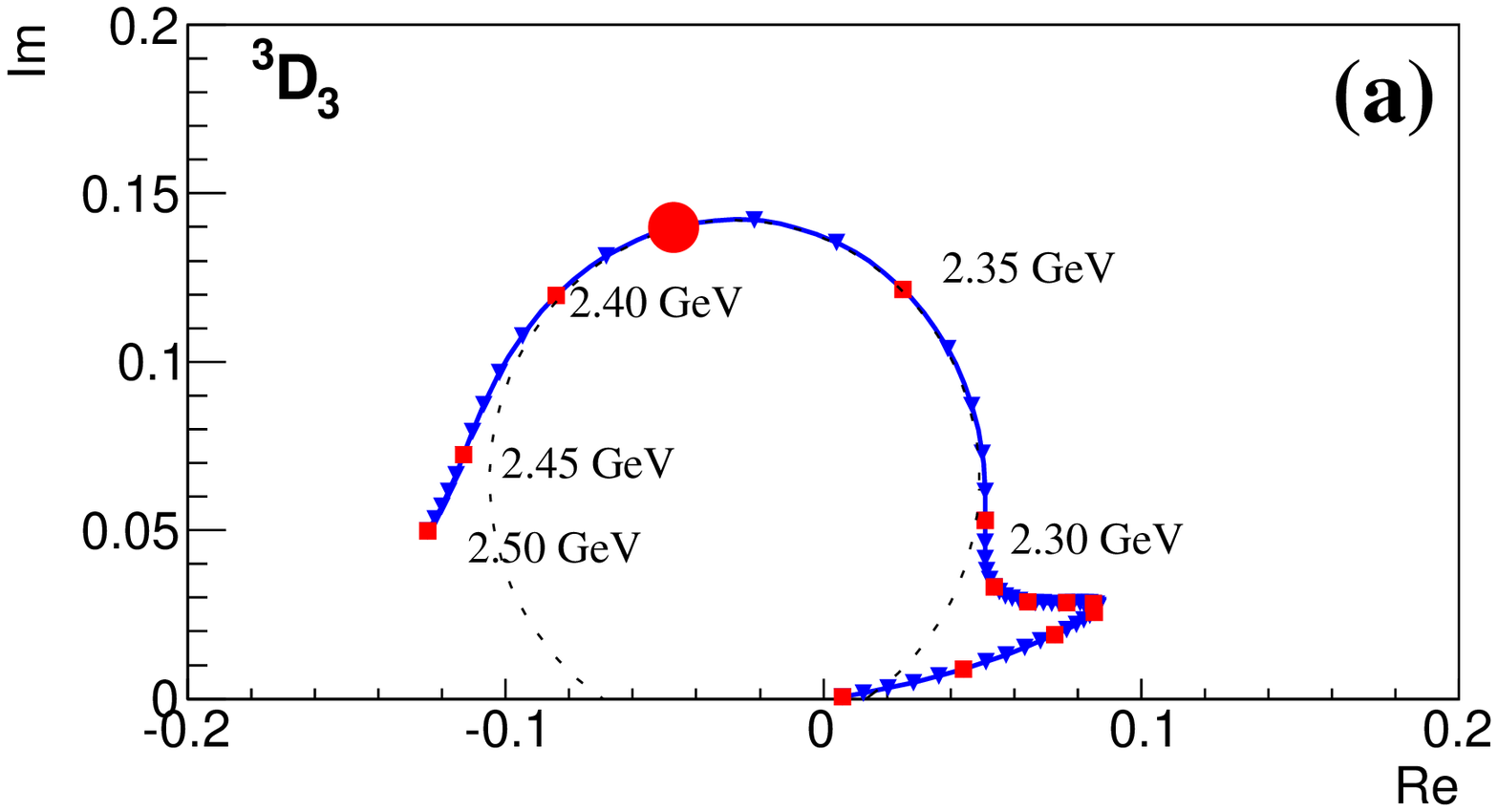} 
\caption{$d^{\ast}$(2380) dibaryon resonance signatures in recent 
WASA-at-COSY Collaboration experiments. Left: from the peak observed in 
the $pn\to d\pi^0\pi^0$ reaction \cite{wasa11}. Right: from the Argand 
diagram of the $^3D_3$ partial wave in $pn$ scattering \cite{wasa14}.} 
\label{fig:WASA} 
\end{center} 
\end{figure} 

The ${\cal D}_{03}$ dibaryon of Table~\ref{tab:dyson} shows up in the $^3D_3$ 
nucleon-nucleon partial wave above the $NN\pi\pi$ threshold owing to the 
coupling between the $I=0$ $^3D_3$ $NN$ channel and the $I=0$ $^7S_3$ $\Delta
\Delta$ channel, i.e. the coupling between the two-body $NN$ channel and 
the four-body $NN\pi\pi$ channel. Indeed its best demonstration is by the 
relatively narrow peak about 80~MeV above the $\pi^0\pi^0$ production 
threshold and 80~MeV below the $\Delta\Delta$ threshold, with $\Gamma_{d^{
\ast}}\approx 70$~MeV, observed in $pn\to d\pi^0\pi^0$ by the WASA-at-COSY 
Collaboration~\cite{wasa11} and shown in Fig.~\ref{fig:WASA}-left.   
The $I=0$ isospin assignment follows from the isospin balance in $pn \to 
d\pi^0\pi^0$, and the $J^P=3^+$ spin-parity assignment follows from the 
measured deuteron angular distribution. The $d^{\ast}$(2380) was also observed 
in $pn\to d\pi^+\pi^-$~\cite{wasa13a}, with cross section consistent with that 
measured in $pn\to d\pi^0\pi^0$, and studied in several $pn\to NN\pi\pi$ 
reactions~\cite{wasa13b,wasa13c,hades15}. Recent measurements of $pn$ 
scattering and analyzing power~\cite{wasa14} have led to the $pn$ $^3D_3$ 
partial-wave Argand diagram shown in Fig.~\ref{fig:WASA}-right, 
supporting the ${\cal D}_{03}$ dibaryon resonance interpretation.  

\begin{table}[hbt] 
\begin{center} 
\caption{${\cal D}_{03}$ mass (in GeV) predicted in several quark-based 
calculations prior to 2008. Wherever calculated, the mass of ${\cal D}_{12}$ 
is also listed.} 
\begin{tabular}{cccccccccc} 
\hline 
${\cal D}_{IS}$ ($BB'$) & \cite{DX64} & \cite{MAS80} & \cite{OY80} & 
\cite{cvetic80} & \cite{MT83} & \cite{Gold89} & \cite{beijing99} & 
\cite{sal02} &~~exp./phen. \\ 
\hline 
${\cal D}_{03}$ ($\Delta\Delta$) &~2.35~&~2.36~&~2.46~&~2.42~&~2.38~& 
$\leq$2.26~&~2.40~&~2.46~& 2.38 \\ 
${\cal D}_{12}$ ($N\Delta$) &~2.16~&~2.36~& -- & -- &~2.36~& -- & -- & 
~2.17~& $\approx$2.15 \\  
\hline 
\end{tabular} 
\label{tab:QM} 
\end{center} 
\end{table} 

The history and state of the art of the ${\cal D}_{03}$ dibaryon, now 
denoted d$^\ast$(2380), were reviewed recently by Clement~\cite{clement17}. 
In particular, its mass was predicted in several quark-based calculations, 
as listed in Table~\ref{tab:QM} in the columns following the symmetry-based 
value predicted first by Dyson and Xuong~\cite{DX64}. Also listed are 
${\cal D}_{12}$ mass values, wherever available from such calculations. 
Remarkably, none of these quark-based predictions managed to reproduce the 
empirical mass values listed in the last column for {\it both} ${\cal D}_{12}$ 
and ${\cal D}_{03}$. More recent quark-based calculations, following the 2008 
first announcement of observing the ${\cal D}_{03}$~\cite{clement08}, are 
discussed below. 

Values of ${\cal D}_{03}$ and ${\cal D}_{30}$ pole positions $W=M-{\rm i}
\Gamma/2$ from our hadronic-model three-body $\pi N\Delta$ Faddeev 
calculations~\cite{GG13,GG14} are listed in Table~\ref{tab:DelDel}. 
The ${\cal D}_{03}$ mass and width values calculated in the Faddeev hadronic 
model version using $r_{\Delta}\approx\,1.3$~fm are remarkably close to the 
experimentally reported ones, whereas the mass evaluated in the model version 
using $r_{\Delta}\approx\,0.9$~fm agrees, perhaps fortuitously so, with that 
derived in the Dyson-Xuong pioneering discussion \cite{DX64}. For smaller 
values of $r_{\Delta}$ one needs to introduce explicit vector-meson and/or 
quark-gluon degrees of freedom which are outside the scope of the present 
model. In contrast, the calculated widths $\Gamma$ are determined primarily 
by the phase space available for decay, displaying little sensitivity to the 
radius $r_{\Delta}$ of the $\pi N$ $P_{33}$ form factor. 

\begin{table}[hbt] 
\begin{center} 
\caption{${\cal D}_{03}$ and ${\cal D}_{30}$ $\Delta\Delta$ dibaryon 
$S$-matrix pole position $W=M-{\rm i}\frac{\Gamma}{2}$ (in MeV), obtained in 
our hadronic model by solving the $\Delta\Delta$ $T$-matrix integral equation 
of Fig.~\ref{fig:piNB}, are listed for two choices of the $\pi N$ $P_{33}$ 
form factor specified by a radius parameter $r_{\Delta}$ (in fm). The last 
two columns list results of post 2008 quark-based RGM calculations with 
hidden-color $\Delta_8\Delta_8$ components. 
} 
\begin{tabular}{ccccccccc} 
\hline 
$\Delta\Delta$ & & \multicolumn{3}{c}{Faddeev (present)} & & 
\multicolumn{3}{c}{Recent quark-based} \\ 
${\cal D}_{IS}$ & & $r_{\Delta}\approx\,1.3$ & & $r_{\Delta}\approx\,0.9$ & & 
Ref.~\cite{wang14} & & Ref.~\cite{dong16}  \\  
\hline 
${\cal D}_{03}$ & & 2383$-{\rm i}$41 & & 2343$-{\rm i}$24 & & 2393$-{\rm i}$75 
& & 2380$-{\rm i}$36 \\ 
${\cal D}_{30}$ & & 2411$-{\rm i}$41 & & 2370$-{\rm i}$22 & & 2440$-{\rm i}
$100 & & -- \\ 
\hline 
\end{tabular} 
\label{tab:DelDel} 
\end{center} 
\end{table} 

The ${\cal D}_{30}$ dibaryon resonance is found in our $\pi N\Delta$ Faddeev 
calculations to lie about 30~MeV above the ${\cal D}_{03}$. These two states 
are degenerate in the limit of equal $D={\cal D}_{12}$ and $D={\cal D}_{21}$ 
isobar propagators in Fig.~\ref{fig:piNB}. Since ${\cal D}_{12}$ was found to 
lie lower than ${\cal D}_{21}$, we expect also ${\cal D}_{03}$ to lie lower 
than ${\cal D}_{30}$ as satisfied in our Faddeev calculations. Moreover, 
here too the difference in their flavor-SU(3) classification will push the 
${\cal D}_{30}$ further apart from the ${\cal D}_{03}$. The ${\cal D}_{30}$ 
has not been observed and only upper limits for its production in $pp\to 
pp\pi^+\pi^+\pi^-\pi^-$ are available~\cite{wasa16}. 

Finally, we briefly discuss the ${\cal D}_{03}$ mass and width values, listed 
in the last two columns of Table~\ref{tab:DelDel}, from two recent quark-based 
resonating-group-method (RGM) calculations~\cite{wang14,dong16} that add 
$\Delta_{\bf 8}\Delta_{\bf 8}$ hidden-color (CC) components to a $\Delta_{\bf 
1}\Delta_{\bf 1}$ cluster. Interestingly, the authors of Ref.~\cite{wang14} 
have just questioned the applicability of admixing CC components in dibaryon 
calculations~\cite{wang17}. The two listed calculations generate mass values 
that are close to the mass of the d$^{\ast}$(2380). The calculated widths, 
however, differ a lot from each other: one calculation generates a width of 
150~MeV~\cite{wang14}, exceeding substantially the reported value $\Gamma_{d^{
\ast}(2380)}$=80$\pm$10~MeV~\cite{wasa14}, the other one generates a width of 
72~MeV~\cite{dong16}, thereby reproducing the d$^{\ast}$(2380) width. While 
the introduction of CC components has moderate effect on the resulting mass 
and width in the chiral version of the first calculation, lowering the mass 
by 20~MeV and the width by 25~MeV, it leads to substantial reduction of the 
width in the second (also chiral) calculation from 133~MeV to 72~MeV. The 
reason is that the dominant CC $\Delta_{\bf 8}\Delta_{\bf 8}$ components, with 
$68\%$ weight~\cite{dong16}, cannot decay through single-fermion transitions 
$\Delta_{\bf 8}\to N_{\bf 1}\pi_{\bf 1}$ to asymptotically free color-singlet 
hadrons. However, as argued in the next section, these quark-based width 
calculations miss important kinematical ingredients that make the width of 
a single compact $\Delta_{\bf 1}\Delta_{\bf 1}$ cluster considerably smaller 
than $\Gamma_{d^{\ast}(2380)}$. The introduction of substantial $\Delta_{
\bf 8}\Delta_{\bf 8}$ components only aggravates the disagreement.

\section{The width of d$^\ast$(2380), small or large?} 
\label{sec:width} 

The width derived for the $d^{\ast}$(2380) dibaryon resonance by the 
WASA-at-COSY Collaboration and the SAID Data-Analysis-Center is $\Gamma_{d^{
\ast}(2380)}$=80$\pm$10~MeV~\cite{wasa14}. It is much smaller than 230~MeV, 
twice the width $\Gamma_{\Delta}\approx 115$~MeV~\cite{SP07,anisovich12} of 
a single free-space $\Delta$, expected naively for a $\Delta\Delta$ quasibound 
configuration. However, considering the reduced phase space, $M_{\Delta}=1232 
\Rightarrow E_{\Delta}=1232-B_{\Delta\Delta}/2$~MeV in a bound-$\Delta$ decay, 
where $B_{\Delta\Delta}=2\times 1232-2380=84$~MeV is the $\Delta\Delta$ 
binding energy, the free-space $\Delta$ width gets reduced to 81~MeV using 
the in-medium single-$\Delta$ width $\Gamma_{\Delta\to N\pi}$ expression 
obtained from the empirical $\Delta$-decay momentum dependence 
\begin{equation}
\Gamma_{\Delta\to N\pi}(q_{\Delta\to N\pi})=\gamma\,
\frac{q^3_{\Delta\to N\pi}}{q_0^2+q^2_{\Delta\to N\pi}},
\label{eq:gamma}
\end{equation}
with $\gamma=0.74$ and $q_0=159$~MeV~\cite{BCS17}. 
Yet, this simple estimate is incomplete since neither of the two $\Delta$s 
is at rest in a deeply bound $\Delta\Delta$ state. To take account of the
$\Delta\Delta$ momentum distribution, we evaluate the bound-$\Delta$ decay
width ${\overline{\Gamma}}_{\Delta\to N\pi}$ by averaging $\Gamma_{\Delta\to
N\pi}(\sqrt{s_{\Delta}})$ over the $\Delta\Delta$ bound-state momentum-space 
distribution, 
\begin{equation}
{\overline{\Gamma}}_{\Delta\to N\pi}\equiv\langle \Psi^{\ast}(p_{\Delta\Delta})
|\Gamma_{\Delta\to N\pi}(\sqrt{s_{\Delta}})|\Psi(p_{\Delta\Delta})\rangle
\approx \Gamma_{\Delta\to N\pi}(\sqrt{{\overline{s}}_{\Delta}}), 
\label{eq:av}
\end{equation}
where $\Psi(p_{\Delta\Delta})$ is the $\Delta\Delta$ momentum-space 
wavefunction and the dependence of $\Gamma_{\Delta\to N\pi}$ on $q_{\Delta\to 
N\pi}$ for on-mass-shell nucleons and pions was replaced by dependence on 
$\sqrt{s_{\Delta}}$. The averaged bound-$\Delta$ invariant energy squared 
${\overline{s}}_{\Delta}$ is defined by
\begin{equation}
{\overline{s}}_{\Delta}=(1232-B_{\Delta\Delta}/2)^2-P_{\Delta\Delta}^2 ,
\label{eq:s}
\end{equation}
in terms of a $\Delta\Delta$ bound-state r.m.s. momentum $P_{\Delta\Delta}
\equiv{\langle p_{\Delta\Delta}^2\rangle}^{1/2}$.

\begin{table}[hbt]
\begin{center}
\caption{Values of $\sqrt{{\overline{s}}_{\Delta}}\,$, of the corresponding 
decay-pion momentum ${\overline{q}}_{\Delta\to N\pi}$ and of ${\overline{
\Gamma}}_{\Delta\to N\pi}$ (\ref{eq:av}), listed as a function of $R_{
\Delta\Delta}$ using $P_{\Delta\Delta}R_{\Delta\Delta}=\frac{3}{2}$ 
in Eq.~(\ref{eq:s}). The last column lists deduced values of ${\overline{
\Gamma}}_{\Delta\Delta\to NN\pi\pi}$, approximating it by $\frac{5}{3}{
\overline{\Gamma}}_{\Delta\to N\pi}$ (see text).} 
\begin{tabular}{ccccc}
\hline
$R_{\Delta\Delta}$ (fm)~~ & ~~$\sqrt{{\overline{s}}_{\Delta}}$ (MeV)~~ & ~~
${\overline{q}}_{\Delta\to N\pi}$ (MeV)~~ & ~~${\overline{\Gamma}}_{\Delta\to
N\pi}$ (MeV)~~ & ~~${\overline{\Gamma}}_{\Delta\Delta\to NN\pi\pi}$ (MeV) \\
\hline
0.6 & 1083 & 38.3  & 1.6  & 2.6  \\
0.7 & 1112 & 96.6  & 19.3 & 32.1 \\
0.8 & 1131 & 122.0 & 33.5 & 55.8 \\
1.0 & 1153 & 147.7 & 50.6 & 84.4  \\
1.5 & 1174 & 170.4 & 67.4 & 112.3  \\
2.0 & 1181 & 177.9 & 73.2 & 122.0  \\
\hline
\end{tabular}
\label{tab:width}
\end{center}
\end{table}

In Table~\ref{tab:width}, taken from my recent work~\cite{gal17}, we list 
values of $\sqrt{{\overline{s}}_{\Delta}}$ and the associated in-medium 
decay-pion momentum ${\overline{q}}_{\Delta\to N\pi}$ for several 
representative values of the r.m.s. radius $R_{\Delta\Delta}\equiv {\langle 
r_{\Delta\Delta}^2\rangle}^{1/2}$ of the bound $\Delta\Delta$ wavefunction, 
using the equality sign in the uncertainty relationship $P_{\Delta\Delta}R_{
\Delta\Delta}\geq 3/2$. Listed also are values of the in-medium 
single-$\Delta$ width ${\overline{\Gamma}}_{\Delta\to N\pi}$ obtained 
from Eq.~(\ref{eq:av}). It is implicitly assumed here that the empirical 
momentum dependence (\ref{eq:gamma}) provides a good approximation also for 
off-mass-shell $\Delta$s. Finally, The last column of the table lists values 
of ${\overline{\Gamma}}_{\Delta\Delta\to NN\pi\pi}$ obtained by multiplying 
${\overline{\Gamma}}_{\Delta\to N\pi}$ by two, for the two $\Delta$s, while 
applying to one of them the isospin projection factor 2/3 introduced in the 
Gal-Garcilazo hadronic model~\cite{GG13,GG14} to obey the quantum statistics 
requirements in the leading final $NN\pi\pi$ decay channels. The large spread 
of ${\overline{\Gamma}}_{\Delta\Delta\to NN\pi\pi}$ width values exhibited in 
the table, all of which are much smaller than the 162~MeV obtained by ignoring 
in Eq.~(\ref{eq:s}) the bound-state momentum distribution, demonstrates the 
importance of momentum-dependent contributions. It is seen that a compact 
d$^{\ast}$(2380) with r.m.s. radius $R_{\Delta\Delta}$ less than 0.8~fm is 
incompatible with the experimental value $\Gamma_{d^{\ast}(2380)}$=80$\pm 
$10~MeV from WASA-at-COSY and SAID even upon adding a non-pionic partial 
width $\Gamma_{\Delta\Delta\to NN}\sim 10$~MeV~\cite{wasa14}. 

\begin{figure}[htb]
\begin{center}
\includegraphics[width=0.48\textwidth]{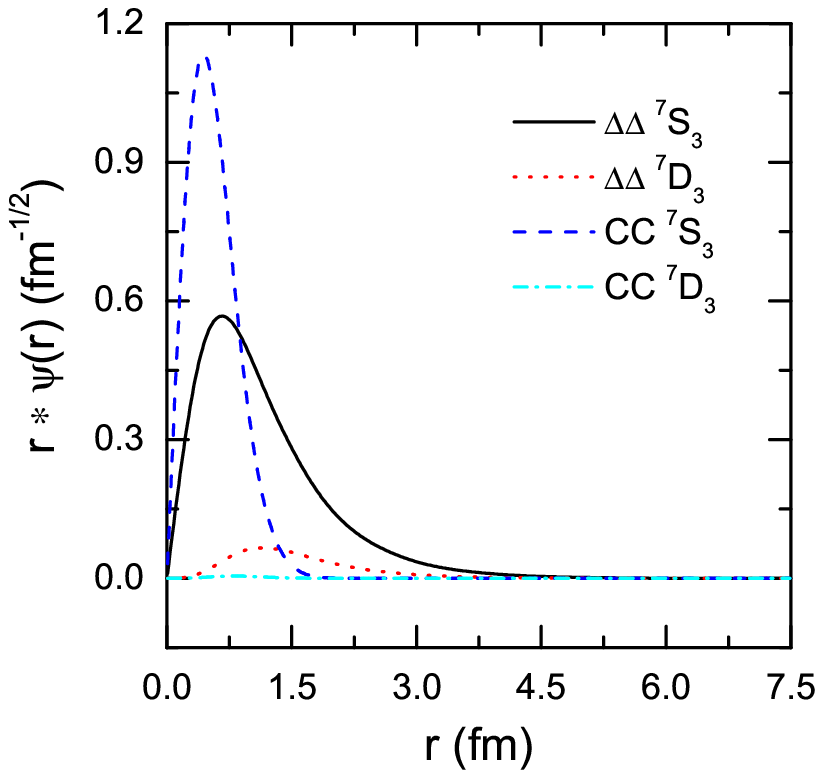}
\includegraphics[width=0.48\textwidth]{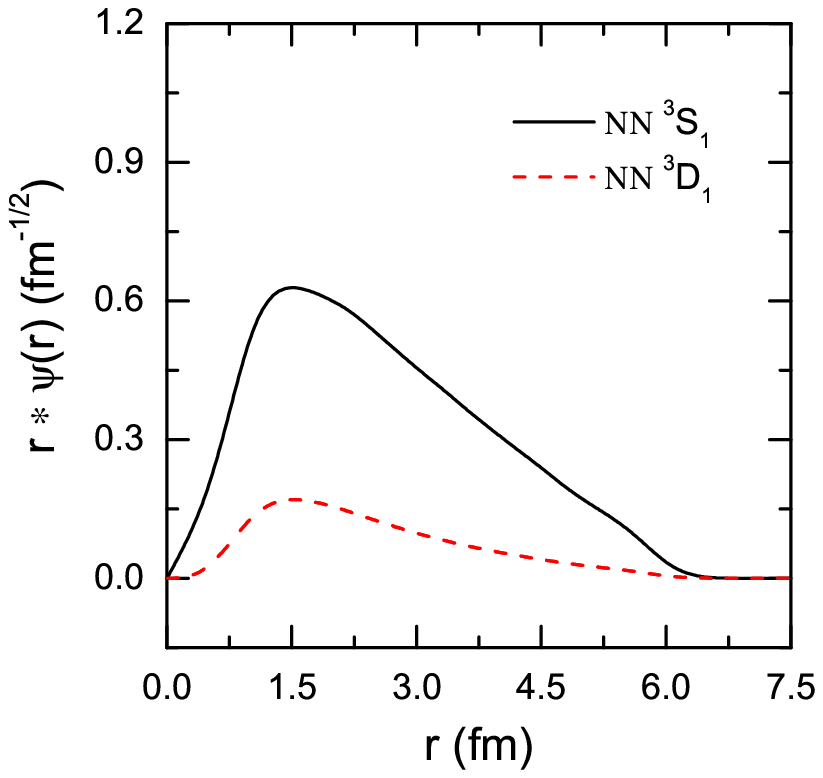}
\caption{d$^{\ast}$(2380) $\Delta\Delta$ wavefunction with r.m.s. 
radius $R_{\Delta\Delta}=0.76$~fm (Left) and deuteron wavefunction 
with r.m.s. radius $R_d\approx 2$~fm (Right) from recent quark-based RGM 
calculations~\cite{dong16,huang15}. Figure adapted from Ref.~\cite{huang15}.} 
\label{fig:wf}
\end{center}
\end{figure}

Fig.~\ref{fig:wf} shows d$^{\ast}$(2380) and d(1876) wavefunctions from 
quark-based RGM calculations~\cite{dong16}. The d$^{\ast}$(2380) appears 
quite squeezed compared to the diffuse deuteron. Its size, $R_{\Delta\Delta}
$=0.76~fm, leads to unacceptably small upper limit of about 47~MeV for the 
d$^{\ast}$(2380) pionic width. This drastic effect of momentum dependence is 
missing in quark-based width calculations dealing with pionic decay modes 
of $\Delta_{\bf 1}\Delta_{\bf 1}$ components, e.g. Ref.~\cite{dong16}. 
Practitioners of quark-based models ought therefore to ask ``what makes 
$\Gamma_{d^{\ast}(2380)}$ so much larger than the width calculated for 
a compact $\Delta\Delta$ dibaryon?" rather than ``what makes $\Gamma_{d^{\ast}
(2380)}$ so much smaller than twice a free-space $\Delta$ width?" 

The preceding discussion of $\Gamma_{d^{\ast}(2380)}$ suggests that the
quark-based model's finding of a tightly bound $\Delta\Delta$ $s$-wave
configuration is in conflict with the observed width. Fortunately, our 
hadronic-model calculations~\cite{GG13,GG14} offer resolution of this 
insufficiency by coupling to the tightly bound and compact $\Delta\Delta$ 
component of the d$^{\ast}$(2380) dibaryon's wavefunction a $\pi N\Delta$ 
resonating component dominated asymptotically by a $p$-wave pion attached 
loosely to the near-threshold $N\Delta$ dibaryon ${\cal D}_{12}$ with size 
about 1.5--2~fm. Formally, one can recouple spins and isospins in this 
$\pi{\cal D}_{12}$ system, so as to assume an extended $\Delta\Delta$-like 
object. This explains why the preceding discussion of $\Gamma_{d^{\ast}\to 
NN\pi\pi}$ in terms of a $\Delta\Delta$ constituent model required a size 
larger than provided by quark-based RGM calculations~\cite{dong16} to 
reconcile with the reported value of $\Gamma_{d^{\ast}(2380)}$. 
We recall that the width calculated in our diffuse-structure $\pi N\Delta$ 
model~\cite{GG13,GG14}, as listed in Table~\ref{tab:DelDel}, is in good 
agreement with the observed width of the d$^{\ast}$(2380) dibaryon resonance. 

\begin{figure}[htb]
\begin{center}
\includegraphics[width=0.9\textwidth]{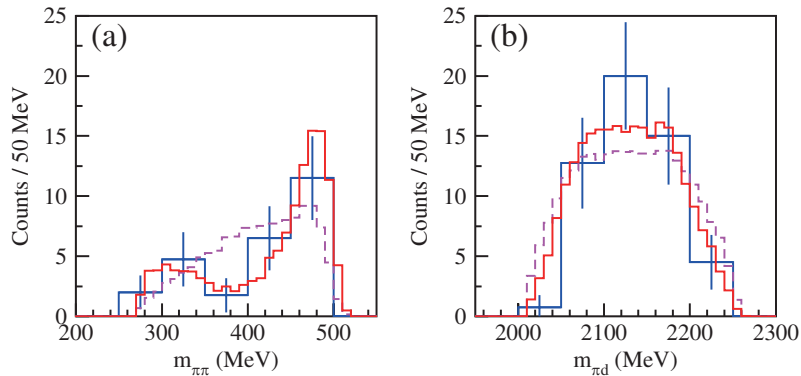} 
\caption{Invariant mass distributions in ELPH experiment~\cite{ELPH17} 
$\gamma d \to d \pi^0 \pi^0$ at $\sqrt{s}=2.39$~GeV.} 
\label{fig:ELPH}
\end{center}
\end{figure}

Support for the role of the $\pi{\cal D}_{12}$ configuration in the decay of 
the d$^{\ast}$(2380) dibaryon resonance is provided by a recent ELPH $\gamma d 
\to d \pi^0 \pi^0$ experiment~\cite{ELPH17} looking for the d$^{\ast}$(2380). 
The cross section data agree with a relativistic Breit-Wigner resonance shape 
with mass of 2370~MeV and width of 68~MeV, but the statistical significance of 
the fit is low, particularly since most of the data are from the energy region 
above the d$^{\ast}$(2380). Invariant mass distributions from this experiment 
at $\sqrt{s}=2.39$~GeV, recorded in Fig.~\ref{fig:ELPH}, are more illuminating. 
The $\pi\pi$ mass distribution shown in (a) suggests a two-bump structure, 
fitted in solid red. The lower bump around 300~MeV is perhaps a manifestation 
of the ABC effect~\cite{ABC60}, already observed in $pn\to d\pi^0\pi^0$ by 
WASA-at-COSY~\cite{wasa11,BCS17} and interpreted in Ref.~\cite{gal17} as due 
to a tightly bound $\Delta\Delta$ decay with reduced $\Delta \to N \pi$ phase 
space. The upper bump in (a) is consistent then with the d$^{\ast}(2380)\to 
\pi {\cal D}_{12}$ decay mode, in agreement with the $\pi d$ mass distribution 
shown in (b) that peaks slightly below the ${\cal D}_{12}$(2150) mass. 

\begin{figure}[htb]
\begin{center}
\includegraphics[width=0.6\textwidth]{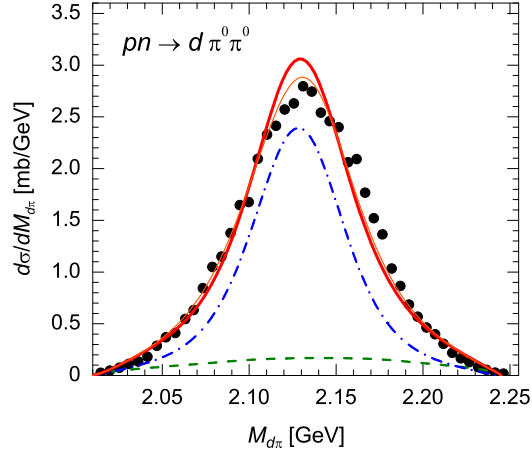}
\caption{The $pn\to d\pi^0\pi^0$ WASA-at-COSY $M_{d\pi}$ invariant-mass 
distribution \cite{wasa11} and, in solid lines, as calculated \cite{PK16} 
for two input parametrizations of ${\cal D}_{12}(2150)$. The dot-dashed 
line gives the $\pi{\cal D}_{12}(2150)$ contribution to the two-body decay 
of the d$^{\ast}$(2380) dibaryon, and the dashed line gives a $\sigma$-meson 
emission contribution. Figure adapted from Ref.~\cite{PK16}.} 
\label{fig:plato}
\end{center}
\end{figure}

Theoretical support for the relevance of the ${\cal D}_{12}(2150)$ $N\Delta$ 
dibaryon to the physics of the d$^{\ast}$(2380) resonance is demonstrated 
in Fig.~\ref{fig:plato}~\cite{PK16} by showing a $d\pi$ invariant-mass 
distribution peaking near the $N\Delta$ threshold as deduced from the $pn\to 
d\pi^0\pi^0$ reaction by which the d$^{\ast}$(2380) was found~\cite{wasa11}. 
However, the peak is shifted to about 20~MeV below the mass of the ${\cal D}_{
12}(2150)$ and the width is smaller by about 40~MeV than the ${\cal D}_{12}
(2150)$ width, agreeing perhaps fortuitously with $\Gamma_{d^{\ast}(2380)}$. 
Both of these features, the peak downward shift and the smaller width, can be 
explained by the asymmetry between the two emitted $\pi^0$ mesons, only one of 
which arises from the $\Delta\to N\pi$ decay within the ${\cal D}_{12}(2150)
$~\cite{PK16} (I am indebted to Heinz Clement for confirming to me this 
explanation). 

\begin{table}[hbt]
\begin{center}
\caption{d$^{\ast}$(2380) decay width branching ratios (BR in percents) 
from Ref.~\cite{gal17} for theory and from Refs.~\cite{BCS15,wasa17} for 
experiment. } 
\begin{tabular}{cccccccccc}
\hline
decay channel & $d\pi^0\pi^0$ & $d\pi^+\pi^-$ & $pn\pi^0\pi^0$ & 
$pn\pi^+\pi^-$ & $pp\pi^-\pi^0$ & $nn\pi^+\pi^0$ & $NN\pi$ & $NN$ & total \\ 
\hline 
BR(th.)  & 11.2 & 20.4 & 11.6 & 25.8 & 4.7 & 4.7 & 8.3 & 13.3 & 100 \\
BR(exp.) & 14$\pm$1 & 23$\pm$2 & 12$\pm$2 & 30$\pm$5 & 6$\pm$1 & 6$\pm$1 & 
$\leq$9 & 
12$\pm$3 & 103 \\ 
\hline
\end{tabular}
\label{tab:BR}
\end{center}
\end{table}

Recalling the $\Delta\Delta$ -- $\pi{\cal D}_{12}$ coupled channel nature of 
the d$^{\ast}$(2380) in our hadronic model~\cite{GG13,GG14}, one may describe 
satisfactorily the d$^{\ast}$(2380) total and partial decay widths in terms 
of an incoherent mixture of these relatively short-ranged ($\Delta\Delta$) 
and long-ranged ($\pi{\cal D}_{12}$) channels. This is demonstrated in 
Table~\ref{tab:BR} where the $NN\pi\pi$ calculated partial widths, totaling 
$\approx$60~MeV, are assigned a weight $\frac{5}{7}$ from $\Delta\Delta$ 
and a weight $\frac{2}{7}$ from $\pi{\cal D}_{12}$. This choice, ensuring 
that the partial decay width $\Gamma_{d^{\ast}\to NN\pi}$ does not exceed 
the upper limit of BR$\leq$9\% determined recently from {\it not} observing 
the single-pion decay branch~\cite{wasa17}, is by no means unique and the 
weights chosen here may be varied to some extent. For more details, see 
Ref.~\cite{gal17}.

\section{Conclusion} 
\label{sec:concl} 

Substantiated by systematic production and decay studies in recent 
WASA-at-COSY experiments \cite{clement17}, the d$^{\ast}$(2380) is the most 
spectacular dibaryon candidate at present. Following its early prediction in 
1964 by Dyson and Xuong~\cite{DX64}, it has been assigned in most theoretical 
works to a $\Delta\Delta$ quasibound state. Given the small width $\Gamma_{d^{
\ast}(2380)}=80\pm 10$~MeV~\cite{wasa14} with respect to twice the width of a 
free-space $\Delta$, $\Gamma_{\Delta}\approx 115$~MeV, its location far from 
thresholds makes it easier to discard a possible underlying threshold effect. 
However, as argued in this review following Ref.~\cite{gal17}, the observed 
small width is much larger than what two {\it deeply bound} $\Delta$ baryons 
can yield upon decay. The d$^{\ast}$(2380) therefore cannot be described in 
terms of a single compact $\Delta\Delta$ state as quark-based calculations 
derive it~\cite{wang14,dong16}. A complementary quasi two-body component is 
provided within a $\pi N\Delta$ three-body hadronic model~\cite{GG13,GG14} 
by the $\pi{\cal D}_{12}$ channel, in which the d$^{\ast}$(2380) resonates. 
The ${\cal D}_{12}$ dibaryon stands here for the $I(J^P)=1(2^+)$ $N\Delta$ 
near-threshold system that might possess a quasibound state $S$-matrix pole. 
It is a loose system of size 1.5--2~fm, as opposed to a compact $\Delta\Delta$ 
component of size 0.5--1~fm. It was also pointed out here, following 
Ref.~\cite{gal17}, how the ABC low-mass enhancement in the $\pi^0\pi^0$ 
invariant mass distribution of the $pn\to d\pi^0\pi^0$ fusion reaction 
at $\sqrt{s}=2.38$~GeV might be associated with a compact $\Delta\Delta$ 
component. The $\pi{\cal D}_{12}$ channel, in contrast, is responsible to the 
higher-mass structure of the $\pi^0\pi^0$ distribution and, furthermore, it 
gives rise to a non-negligible $d^{\ast}\to NN\pi$ single-pion decay branch, 
considerably higher than that obtained for a quark-based purely $\Delta\Delta$ 
configuration~\cite{dong17}, but consistently with the upper limit of $\leq
$9\% determined recently by the WASA-at-COSY Collaboration~\cite{wasa17}. 
A precise measurement of this decay width and BR will provide a valuable 
constraint on the $\pi{\cal D}_{12}$--$\Delta\Delta$ mixing parameter. 

We end with a brief discussion of possible 6q admixtures in the essentially 
hadronic wavefunction of the d$^{\ast}$(2380) dibaryon resonance. For this 
we refer to the recent 6q non-strange dibaryon variational calculation in 
Ref.~\cite{PPL15} which depending on the assumed confinement potential 
generates a $^3S_1$ 6q dibaryon about 550 to 700~MeV above the deuteron, 
and a $^7S_3$ 6q dibaryon about 230 to 350~MeV above the d$^{\ast}$(2380). 
Taking a typical 20~MeV potential matrix element from deuteron structure 
calculations and 600~MeV for the energy separation between the deuteron and 
the $^3S_1$ 6q dibaryon, one finds admixture amplitude of order 0.03 and 
hence 6q admixture probability of order 0.001 which is compatible with that 
discussed recently by Miller~\cite{miller14}. Using the same 20~MeV potential 
matrix element for the $\Delta\Delta$ dibaryon candidate and 300~MeV for the 
energy separation between the d$^{\ast}$(2380) and the $^7S_3$ 6q dibaryon, 
one finds twice as large admixture amplitude and hence four times larger 6q 
admixture probability in the d$^{\ast}$(2380), altogether smaller than 1\%. 
These order-of-magnitude estimates demonstrate that long-range hadronic and 
short-range quark degrees of freedom hardly mix also for $\Delta\Delta$ 
configurations, and that the d$^{\ast}$(2380) is extremely far from a pure 6q 
configuration. This conclusion is at odds with the conjecture made recently 
by Bashkanov, Brodsky and Clement~\cite{BBC13} that 6q CC components dominate 
the wavefunctions of the $\Delta\Delta$ dibaryon candidates ${\cal D}_{03}$, 
identified with the observed d$^{\ast}$(2380), and ${\cal D}_{30}$. 
Unfortunately, most of the quark-based calculations discussed in the present 
work combine quark-model input with hadronic-exchange model input in a loose 
way which discards their predictive power.

\begin{acknowledgement} 

I'm indebted to the organizers of the Frontiers of Science symposium in memory 
of Walter Greiner, held at FIAS, Frankfurt, June 2017, particularly to Horst 
St\"{o}cker, for inviting me to participate in this special event and for 
supporting my trip. Special thanks are due to Humberto Garcilazo, together 
with whom the concept of pion assisted dibaryons was conceived, and also due 
to Heinz Clement for many stimulating exchanges on the physics of dibaryons 
and Jerry Miller for instructive discussions on 6q contributions to dibaryons. 

\end{acknowledgement}

\end{document}